\documentclass[manuscript]{aastex631}   

\usepackage{txfonts}
\usepackage{latexsym,bm}
\usepackage{algorithm, algorithmic}
\newcommand{\txt}[1]{\mathrm{#1}}

\usepackage{stmaryrd}
\usepackage{trimclip}

\makeatletter
\DeclareRobustCommand{\shortto}{%
  \mathrel{\mathpalette\short@to\relax}%
}

\newcommand{\short@to}[2]{%
  \mkern2mu
  \clipbox{{.5\width} 0 0 0}{$\m@th#1\vphantom{+}{\shortrightarrow}$}%
  }
\makeatother

\usepackage{multirow}
\newcommand{\qinemail}{qingang@hit.edu.cn}
\newcommand{\hit}{School of Science, Harbin Institute of Technology, Shenzhen, 
518055, People’s Republic of China}
\newcommand{\hitqin}{\hit; \qinemail}
\newcommand{\szlab}{Shenzhen Key Laboratory of Numerical Prediction for Space Storm,
Harbin Institute of Technology, Shenzhen, 518055, People’s Republic of China}

\shorttitle{Perpendicular Diffusion Effect of CR Anisotropy}
\shortauthors{Meng and Qin}

\turnoffedit

\begin{document}

%\title{A Simple Dipole Anisotropy Amplitude Model}
\title{Perpendicular Diffusion Effect of Cosmic Ray Dipole Anisotropy} 
%Considering B/C Observations}

\correspondingauthor{G. Qin}
\email{\qinemail}

\author{H. R. Meng}
\affiliation{\hitqin}

\author[0000-0002-3437-3716]{G. Qin}
%\altaffiliation{Author of correspondence.}
\affiliation{\hitqin}
\affiliation{\szlab}

%%%%%%%%%%%%%%%%%%%%%%%%% Section O %%%%%%%%%%%%%%%%%%%%%%%%%%%%%%%%%%%%%%%%%%%%%%%%%

\begin{abstract}

Anisotropy is very important to understand cosmic ray (CR) source and 
interstellar environment. The theoretical explanation of cosmic rays anisotropy
 from experiments remains challenging and even puzzling for a long time. In 
this paper, by following the ideas of Jokipii 2007, we use a simple analytical
model to study the CR dipole anisotropy amplitude, considering that CRs diffuse only
in perpendicular direction, with the ratio between the secondary and primary 
cosmic rays omnidirectional particle distribution function as an input. We make
 power law fitting of the observed B/C ratio and use it as the input of the 
anisotropy model. We show that the modeling results can roughly describe the 
general trend of the observational data in energy range from $6\times 10^1$ to 
$3\times 10^{11}$ GeV. It is suggested that the perpendicular diffusion may
play a significant role in CR anisotropy in the wide energy range.

% In this work, we follow ideas of Jokipii 2007 to obtain a model of the cosmic ray 
% (CR) dipole anisotropy amplitude, $\delta$. We use the simple and effective leaky-box 
% model to describe the propagation of CRs, considering the galactic disk radius much 
% larger than the half thickness, with a surrounding galactic halo. We assume the 
% dipole anisotropy amplitude characterized by the angular-dependent CR intensity, considering the 
% high-energy CRs slowly escaping from the Galaxy with perpendicular diffusion, in the 
% approximation of the Fick's diffusion law. A simple CR dipole anisotropy amplitude 
% model is obtained with the ratio between the secondary and primary CRs 
% omnidirectional particle distribution function as an input, in additional to several 
% basic galactic parameters. We fit the observed B/C ratio with a power law as the 
% input for the dipole anisotropy amplitude, the modeling results of $\delta$ is shown 
% to have the similar increasing trend with a power law as the observational data in 
% the whole energy range from $6\times 10^1$ to $3\times 10^{11}$ GeV.

\end{abstract}

%\keywords{}
%\begin{document}

%%%%%%%%%%%%%%%%%%%%%%%%% Section I %%%%%%%%%%%%%%%%%%%%%%%%%%%%%%%%%%%%%%%%%%%%%%%%%
\section{Introduction}
\label{sec:intro}
The cosmic rays (CRs) are primarily composed of charged nuclei that undergo 
diffusion by the galactic magnetic field with turbulence, a phenomenon known as the 
random walk of CRs \citep{Jokipii66}. Over the extended period of the propagation 
from their sources to the heliosphere, CRs experience significant scatterings in 
the interstellar media, losing directional information from the sources and becoming 
highly isotropic. However, as early as in the 1930s, there were signs of the subtle 
CR anisotropy even with significant systematic and statistical uncertainty in data 
\citep{Wollan39}. The confirmation of weak anisotropy was not possible before the 
emergence of large underground muon detectors and extended air shower arrays in the 
1950s, which allowed people to obtain a large amount of reliable data 
\citep{DiSciascioAIuppa14}. With the rapid development of observation technology, 
precise CR data have been obtained with a wide energy range spanning from GeV to 
nearly ZeV. One has reached the following conclusions based on years of accumulated 
observation data \citep{Deligny16}. Generally, the anisotropy of particles increases 
with energy. Specifically, particles at the TeV energy level exhibit an anisotropy 
of approximately $10^{-3}-10^{-4}$, with a local decrease observed at $100$ TeV. 
Meanwhile, particles at the EeV energy level demonstrate an anisotropy of 
approximately $10^{-2}$, with a local decrease observed at a few EeV.

The anisotropy is a powerful tool for investigating CR sources and interstellar 
magnetic field information, so it has attracted much attention from researchers. 
\citet{ComptonAGetting35} derived the Compton-Getting anisotropy based on the 
observer's motion relative to the rest frame of the isotropic CRs. Compton-Getting 
anisotropy is independent of energy and clearly inconsistent with observations. 
To account for the faint anisotropy observed in CRs after the diffusion process, 
\citet{BerezinskiiEA90} proposed a simple isotropic diffusion model, which, however, 
is inconsistent with observations in the TeV-PeV energy range. Additionally, there 
are three parameters in this simple model that still require investigation: the 
diffusion coefficient, the CR density, and its gradient. To address this problem, 
some researchers \citep[e.g.,][] {JonesEA01, ErlykinAWolfendale06, BlasiAAmato12, 
PohlAEichler13} have augmented the standard isotropic diffusion model by taking into 
account properties of CR sources. But the predicted dipole anisotropy amplitude over 
TeV energy still significantly exceeds observations. Others have put forth more 
innovative perspectives, which have only partially reduced the disparity between 
results and observations. For instance, \citet{Zirakashvili05} provided a method to 
reduce the diffusion coefficient in the solar magnetic field; \citet{EvoliEA12} 
considered the diffusion coefficient varying with changes in the turbulent sources 
of the interstellar magnetic field in space; \citet{MertschAFunk15} studied the 
discrepancy between the regular magnetic field and the CR density gradient; 
Particularly, both \citet{QiaoEA19} and \citet{ZhangEA2} investigated the CR  
anisotropies of different mass groups on the setting of many free parameters to 
obtain the dipole anisotropy, which is almost consistent with observations. The 
former employed a spatially-dependent propagation model while the later utilized a 
three-component model. However, to confirm these interesting speculations one need 
further observations. Overall, these new models, in order to address the parameters 
of the standard isotropic diffusion model, often adopt additional parameters and 
employ more complicated numerical methods. Furthermore, several CR anisotropy models 
different from the standard diffusion theory have also been proposed, to provide 
quantitative results regarded as an order of magnitude estimation, as the 
uncertainties in the parameters used in the calculations. For instance, considering 
the galactic CR streaming and the galactic halo magnetic structures, \citet{QuEA12} 
created a global galactic CR stream model; \citet{ZhangEA14} proposed a local origin 
model and suggested the ultimate source of TeV CR anisotropy seen at Earth is the 
inhomogeneity of CR distribution function in the local interstellar medium, which 
can give rise to three types of anisotropies; \citet{SchlickeiserEA19} derived two 
types of CR anisotropy from the Fokker-Planck equation based on quasilinear 
diffusion because of the resonant interactions with slab waves; \citet{ZhangEA22} 
discussed the inertial anisotropy and shear anisotropy generated by non-uniform 
convection in the Bhatnagar–Gross–Krook approximation. 

In the workshop presentation, \citet{Jokipii07} suggested that CR escape time 
$\tau_\txt{e}$, the time taken by CRs to escape in the direction perpendicular to 
the galactic disk, can be propotional to the ratio between the secondary and primary 
nuclei omnidirectional particle distribution function, $f_\txt{s}/f_\txt{p} \propto 
\tau_\txt{e}$, and that $\tau_\txt{e}$ is approximately $20$ Myr at GeV energies 
from observations. In addition, \citet{Jokipii07} asserted that CR dipole anisotropy 
amplitude, $\delta$, can be obtained from the diffusion approximation with the 
leaky-box model, $\delta\approx 3L/(\tau_\txt{e} c)$, where $L$ is the macroscopic 
scale, so $\delta\approx 10^{-4}$ for GeV energies. In this work, following 
\citet{Jokipii07}, we show a simple analytical model of the CR dipole anisotropy
amplitude based on the leaky-box model with the observations of B/C ratio as the input. 
The derivation of the dipole anisotropy amplitude model is given in 
Section~\ref{sec:appraoch}. The modeling results and their comparison with the 
observations in the energy range from tens of GeV to hundreds of EeV are presented 
in Section~\ref{sec:results}. Section~\ref{sec:discussion} is devoted to conclusion 
and discussion.

%%%%%%%%%%%%%%%%%%%%%%%%% Section II %%%%%%%%%%%%%%%%%%%%%%%%%%%%%%%%%%%%%%%%%%%%%%%%
\section{Approach and Method}
\label{sec:appraoch}

As suggested by \citet{Jokipii07}, we use the simple and effective leaky-box model 
to describe the propagation of CRs \citep{Cesarsky80}. We assume that the galactic 
disk has a radius $R$ which is much larger than the half thickness $h$ in $z$ 
direction, $R\gg h$. It is also assumed that $h=100$ pc and the gas number density 
is $n_\txt{d}$. In addition, a galactic halo, with a half thickness $H$ and mean gas 
number density $\bar{n}$, is supposed to be around the galactic disk 
\citep{Blasi13}. The magnetic field is generally perpendicular to the 
z-direction. The CRs are considered to escape only in the $z$ (perpendicular) 
direction with escape time $\tau_\txt{e}$ since the perpendicular escape time can be 
assumed much smaller than the parallel one \citep{EvoliEA12}. Therefore, the 
leaky-box model assumes that the halo is a closed-box defined by a boundary with a 
finite escape probability in $z$ direction, in which CRs are confined for some time, 
$\tau_\txt{e}$, before escaping into the intergalactic space with perpendicular 
diffusion. 

It is known that the perpendicular diffusion coefficient of CRs, $D_\perp(E)$, is a 
function of energy per nucleon $E$, but it is nearly independent of the CR species 
\citep{Jokipii66, MatthaeusEA03}. We can use the escape (also confinement) time that 
is defined by 
\begin{equation}
\tau_{\txt{e}}=\frac{{H^{2}}}{{2 D_\perp(E)}}
\label{eq:taue}
\end{equation}
to characterize the finite escape probability. The escape time $\tau_{\txt{e}}$ is 
thus supposed to be independent of the CR species. When high-energy CRs slowly 
escape from the Galaxy, one expects to detect anisotropies on large angular scales. 
For simplicity let us consider the case of the dipole anisotropy amplitude 
characterized by the angular-dependent CR intensity. In the approximation of the 
Fick's diffusion law, the CR dipole anisotropy amplitude $\delta$ can be shown as 
\citep{BerezinskiiEA90}
\begin{equation}
    \delta = \frac{{3 D_\perp(E)}}{{n v}}\left|\frac{\partial n}{\partial z}\right|,
    \label{eq:A1}
\end{equation}
where $n$ is the CR number density, and CR speed $v=c$ for high energy particles. 
Note that $z$ is supposedly in the perpendicular direction. We assume that the 
galactic halo with half thickness $H$ is the characteristic scale of z-axis 
direction, so that the CR density gradient $\nabla n \sim \frac{{\partial n}} 
{{\partial z}} \sim \frac{n}{H}$. Considering Equations~(\ref{eq:taue}) 
and~(\ref{eq:A1}), the CR dipole anisotropy amplitude $\delta $ can be rewritten as 
\begin{equation}
    \delta = \frac{{3 H}}{{2 c \tau_{\txt{e}}}}.
    \label{eq:A2}
\end{equation}
\citet{Jokipii07} suggested similar results considering the macroscopic scale $L$.

In order to estimate the dipole anisotropy amplitude $\delta$, we need to evaluate 
the CR escape time $\tau_{\txt{e}}$. Since the CR sources are assumed to be located 
in the galactic disk, the mean density can be estimated as \citep{Blasi13}
\begin{equation}
\bar{n} = n_{\txt{d}} \frac{h}{H}.
\label{eq:nbar}
\end{equation}

Under the stationary condition, \citet{GloecklerAJokipii69} used the leaky-box model 
with the equilibrium equation to describe propagation of CRs. In this work, 
neglecting other processes such as convection, re-acceleration, and fragmentation, 
one can obtain a simplified leaky-box model,
\begin{equation}
    \frac{{\partial f}}{{\partial t}} \approx 0 
    \approx - \frac{f}{\tau_{\txt{e}}} + Q,
    \label{eq:LBM}
\end{equation}
where $f$ is the omnidirectional particle distribution function, $t$ is the time, 
and $Q$ is the source of particles. In the Galaxy, primary CRs with stable nuclei 
such as H, He, C, O, and Fe can undergo inelastic collisions (spallation) with the 
interstellar medium during propagation, producing secondary CRs with lighter nuclei 
such as Li, Be, B, and F. For the secondaries, Equation~(\ref{eq:LBM}) can be 
written as
\begin{equation}
    f_{\txt{s}} = Q_{\txt{s}} \tau_{\txt{es}}.
    \label{eq:secondary}
\end{equation}
As noted with Equation~(\ref{eq:taue}), the escape time is independent of CR 
species, so the escape time of secondary CRs $\tau_\txt{es}=\tau_\txt{e}$. In 
addition, the source term can be given by \citep[e.g.,][]{ReevesEA70, 
PtuskinASoutoul90, BerezhkoEA03}
\begin{equation}
 Q_{\txt{s}} = \bar{n} v \sigma_{\txt{p} \shortto \txt{s}} f_{\txt{p}}, 
 \label{eq:Qs}
\end{equation}
where $v=c$ is the speed of particles, $\sigma_{\txt{p} \shortto \txt{s}}$ is the 
primary spallation cross-section, and $f_{\txt{p}}$ is the particle distribution 
function of primary CRs. Considering 
Equations~(\ref{eq:nbar}),~(\ref{eq:secondary}), and~(\ref{eq:Qs}), the escape time 
$\tau_{\txt{e}}$ becomes
\begin{equation}
    \tau_{\txt{e}} = \frac{f_{\txt{s}}}{f_{\txt{p}}} 
        \frac{1}{{c n_{\txt{d}}\sigma_{\txt{p} \shortto \txt{s}} }} \frac{H}{h}.
    \label{eq:ET}
\end{equation}
Note that \citet{Jokipii07} also suggested that $\frac{f_{\txt{s}}}{f_{\txt{p}}}$ is 
propotional to the escape time.

According to Equations~(\ref{eq:A2}) and~(\ref{eq:ET}), the dipole anisotropy 
amplitude becomes
\begin{equation}
    \delta = \frac{3}{2} h n_{\txt{d}} \sigma_{\txt{p} \shortto \txt{s}} 
             \left(\frac{f_{\txt{s}}}{f_{\txt{p}}}\right)^{-1}.
    \label{eq:A3_0}
\end{equation}
Here, we need the secondary-to-primary ratio of CRs, 
$\frac{f_{\txt{s}}}{f_{\txt{p}}}$, and the corresponding cross-section, 
$\sigma_{\txt{p} \shortto \txt{s}}$. We choose the observation of the B and C as the 
secondary and primary CRs, respectively, since B/C ratio is found to be the most abundant 
and accurate among all the available secondary-to-primary ratios. Hereafter, we use 
B/C ratio to represent $\frac{f_\txt{s}}{f_\txt{p}}$. Correspondingly, the spallation 
cross-section, $\sigma_{\txt{C} \shortto \txt{B}}$, is used for $\sigma_{\txt{p} 
\shortto \txt{s}}$. Finally, the Equations~(\ref{eq:A3_0}) becomes
\begin{equation}
    \delta = \frac{3}{2} h n_{\txt{d}} \sigma_{\txt{C} \shortto \txt{B}}  
             \left(\frac{B}{C}\right)^{-1}.
    \label{eq:A3}
\end{equation}
It is noted that this is an analytical model.
%%%%%%%%%%%%%%%%%%%%%%%%% Section III %%%%%%%%%%%%%%%%%%%%%%%%%%%%%%%%%%%%%%%%%%%%%%%
\section{Modeling Results}
\label{sec:results}
In the following, we get the numerical results of the CR dipole anisotropy amplitude 
with Equation~(\ref{eq:A3}). According to the current generation of CR experiments 
\citep{GénoliniEA18}, $\sigma_{\txt{C} \shortto \txt{B}}$ is set to $30.0$ mbarn. 
The typical gas density of the galactic disk that consists of $90\%$ H and $10\%$ 
He, $n_{\txt{d}}$, is set to $1$ atom/cm$^{3}$ \citep{Blasi13}. The latest 
observational CR data of the B/C ratio, which have reached $4\times 10^3$ GeV/n, are 
obtained from the CRDB Website (\url{https://lpsc.in2p3.fr/crdb}). In addition, from 
the same website we also get the latest observational dipole anisotropy amplitude, 
which are in the total energy $E_\txt{tot}$ range from $6\times 10^1$ to $3\times 
10^{11}$ GeV. Note that, the observations of B/C ratio are as a function of energy per 
nucleon $E$, so the model of the CR dipole anisotropy amplitude with 
Equation~(\ref{eq:A3}) shows a function of energy per nucleon $E$, instead of total 
energy $E_\txt{tot}$. However, the observed CRs are mostly protons, so we ignore the 
difference between the total energy $E_\txt{tot}$ and the energy per nucleon $E$ for 
observations and model results, respectively, considering the CR dipole anisotropy 
amplitude $\delta$.

\begin{figure}
\epsscale{1} \plotone{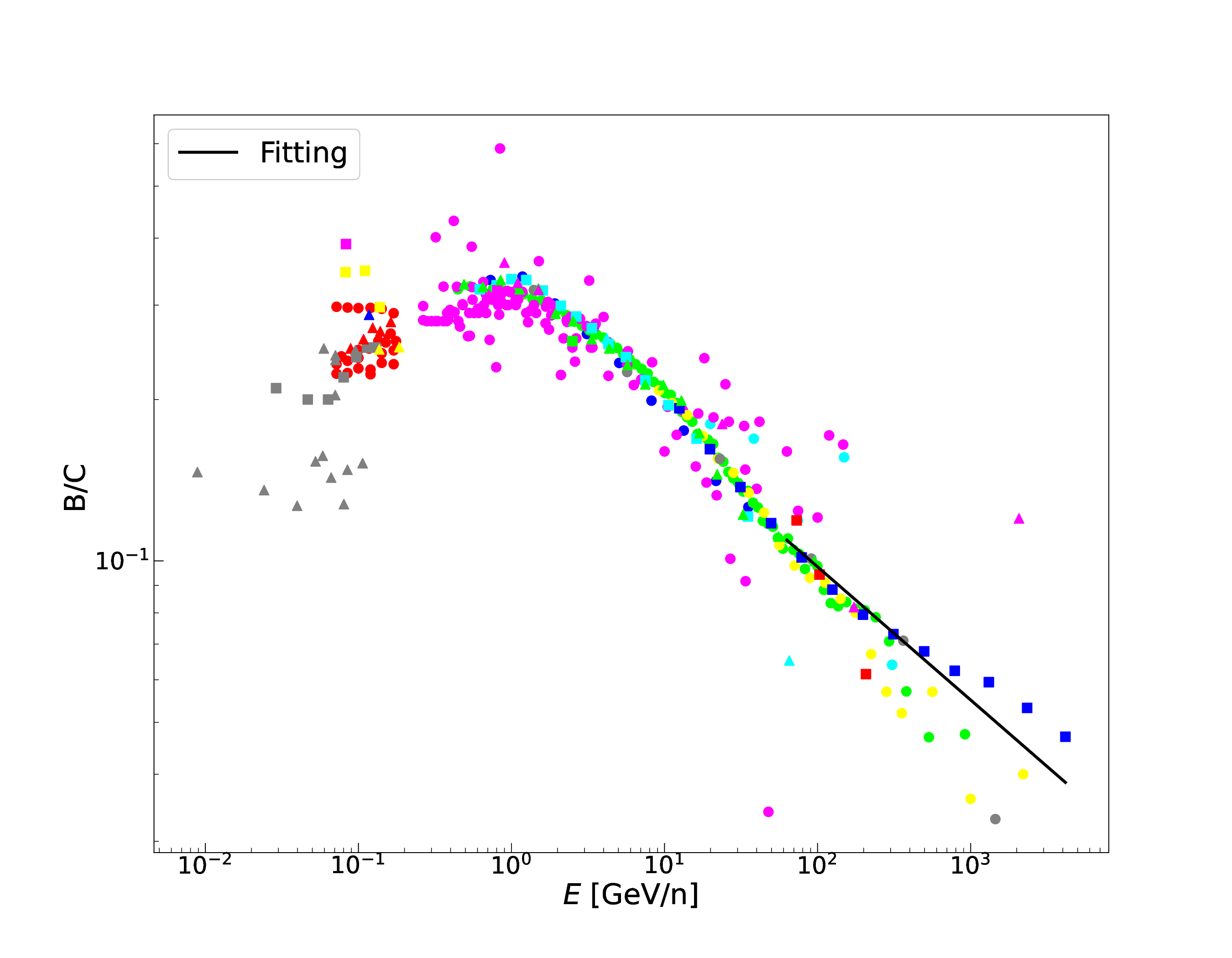}
\caption{The B/C ratio as a function of energy per nucleon, $E$. Colored symbols
indicate the observational data with different experiments. Colored symbols
indicate the observational data with different experiments, as shown in Table~\ref{tab:B/C}. Black solid line shows the 
fitting results of the observational data for energies above $60$ GeV/n.} 
\label{fig:B/C}
\end{figure}

Figure~\ref{fig:B/C} shows all the observational data available for the B/C ratio, 
as a function of energy per nucleon, $E$, with various colors and symbols indicating 
the corresponding experiments as shown in Table~\ref{tab:B/C}. It is shown that 
there is a peak at about $1$ GeV/n for B/C ratio, and a power law may be fitted in high 
energy \citep{Blasi13}. One can suggest that CRs are not significantly influenced by 
solar modulation if their energies are above $10$ GeV/n 
\citep[e.g.,][]{StraussAPotgieter14, QinAShen17}. In this work, with energy range 
from $6\times 10^1$ to $4\times 10^3$ GeV/n we fit B/C ratio with a power law
\begin{equation}
    \left(\frac{B}{C}\right) = k \left(\frac{E}{E_0}\right)^{-\psi},
    \label{eq:fitting}
\end{equation}
where $E_0 = 1$ GeV/n, and $k$ and $\psi$ are fitting parameters. The solid line in 
Figure~\ref{fig:B/C} shows the fitting results with parameters $k \approx 0.31$ and 
$\psi \approx 0.25$. Here, we set the lower limit of the fitting energy range to be 
$6\times 10^1$ GeV/n to align it with the lower limit of the energy range for 
anisotropy observations.

\begin{figure}
\epsscale{1} \plotone{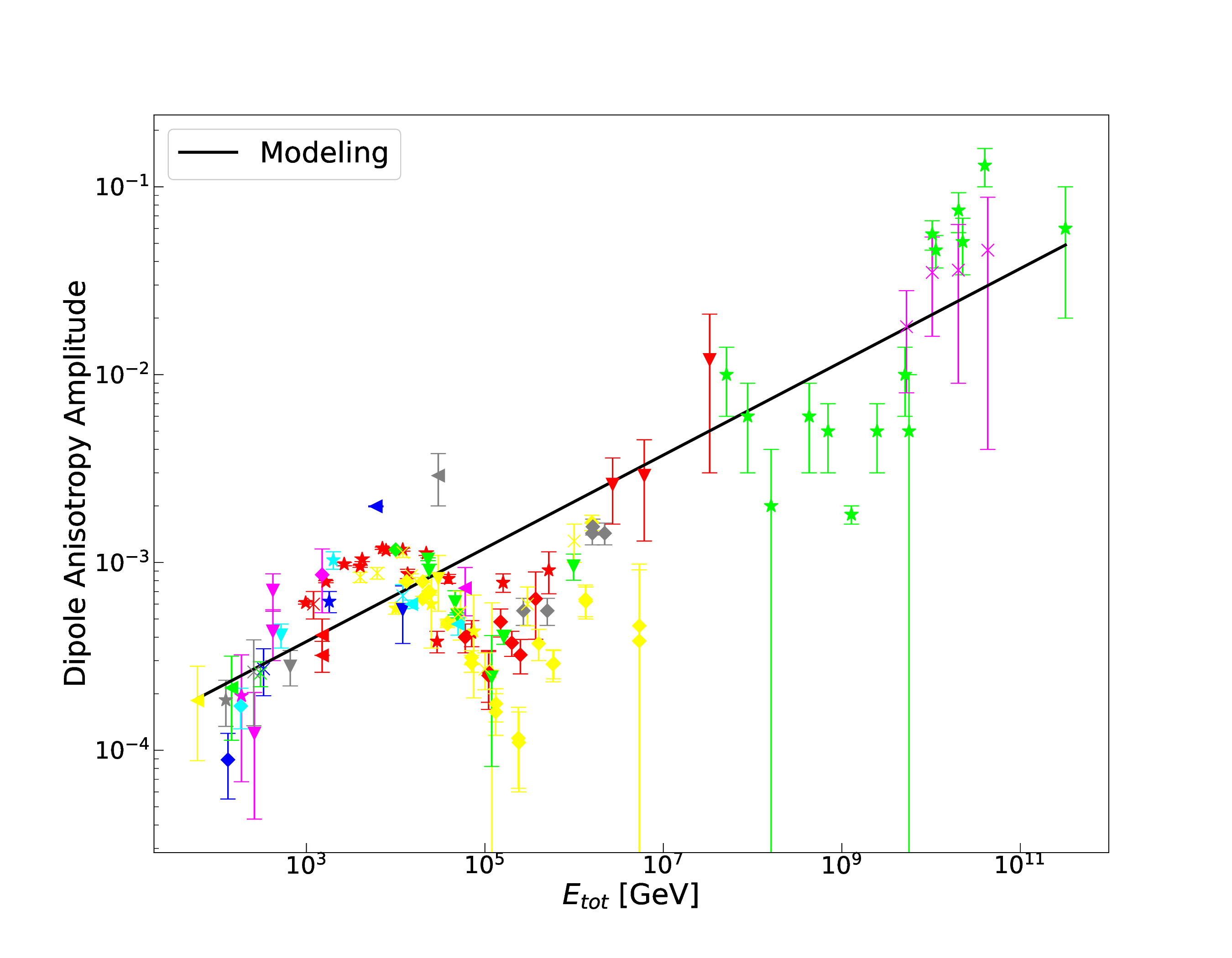}
\caption{Dipole anisotropy amplitude as a function of total energy, $E_\txt{tot}$.
Colored symbols indicate the observational data with different experiments. Colored 
symbols indicate the observational data with different experiments, as shown in
Table~\ref{tab:anisotropy}. Error 
bars show the total experimental errors. Black solid line shows the modeling results.}
\label{fig:anisotropy}
\end{figure}

Figure~\ref{fig:anisotropy} shows the CR dipole anisotropy amplitude $\delta$ as a 
function of total energy $E_\txt{tot}$. The colored symbols indicate the 
observational data for different experiments, with error bars showing the total 
experimental errors of the data. The colored symbols indicate the observational data 
for different experiments, with error bars showing the total experimental errors of 
the data, as shown in Table~\ref{tab:anisotropy}. The black solid line indicates the 
modeling results from Equation~(\ref{eq:A3}), with B/C ratio from the fitting results
Equation~(\ref{eq:fitting}). Here, we ignore the difference between the total energy 
and energy per nucleon for CR anisotropy. It is noted that the fitting of the 
observational data of B/C ratio are in the energy range $6\times 10^1-4\times 10^3$ GeV/n, 
but to obtain the modeling results of the dipole anisotropy amplitude, the fitting 
results of B/C ratio are used in the energy range $6\times 10^1-3\times 10^{11}$ GeV. From 
Figure~\ref{fig:anisotropy} we can see that, generally, the observational data and 
modeling results of the dipole anisotropy amplitude have the similar increasing 
trend with a power law in the whole energy range, $6\times 10^1-3\times 10^{11}$ 
GeV, apart from the two declining regions around $10^5$ GeV and $10^9$ GeV in 
observations. Note that the data uncertainty in the ultra-high energy range is more 
significant. 
%%%%%%%%%%%%%%%%%%%%%%%%%%% Section IV %%%%%%%%%%%%%%%%%%%%%%%%%%%%%%%%%%%%%%%%%%%%%%
\section{Conclusions and Discussion}
\label{sec:discussion}

In this work, we follow \citet{Jokipii07} to derive a model of the CR dipole 
anisotropy amplitude. We use the simple and effective leaky-box model to describe 
the propagation of CRs. In the model, the galactic disk is assumed to have a radius 
which is much larger than the half thickness in $z$ direction. In addition, a 
galactic halo is supposed to be around the galactic disk. The CRs are considered to 
escape into the intergalactic space in the $z$ (perpendicular) direction with escape 
time $\tau_\txt{e}$ with perpendicular diffusion, assuming the perpendicular escape 
time much smaller than the parallel one. In addition, we assume the dipole 
anisotropy amplitude characterized by the angular-dependent CR intensity, 
considering the high-energy CRs slowly escaping from the Galaxy, in the 
approximation of the Fick's diffusion law. To estimate the dipole anisotropy 
amplitude, we also evaluate the CRs escape time using the simplified leaky-box model 
considering the secondaries of CRs for the inelastic collisions of the primaries. 
Note that the escape time is assumed to be independent of CR species. Thus, the CR 
dipole anisotropy amplitude model is obtained with the ratio between the secondary 
and primary CRs omnidirectional particle distribution function as an input. To get 
the modeling results of the CR dipole anisotropy amplitude, we fit the B/C ratio 
with a power law with energy range from $6\times 10^1$ to $4\times 10^3$ GeV/n. 
Using the fitting results of the power law B/C ratio, we get the modeling results of the 
dipole anisotropy amplitude. It is shown that the observational data and modeling 
results of the dipole anisotropy amplitude have the similar increasing trend with a 
power law in the whole energy range from $6\times 10^1$ to $3\times 10^{11}$ GeV.
Therefore, we suggest that the perpendicular diffusion may
play a significant role in CR anisotropy in the energy range.

Typically, researchers have employed more complex models that incorporate 
various effects when studying cosmic-ray anisotropy, and their results can be 
well-fitted to the observational data of cosmic rays, even to the extent of 
reproducing some of the finer details. However, their models and the 
computation are often complex, which also requires various tuning factors to 
improve the model outcomes. The model we adopt is relatively simple, hence the 
agreement between our model predictions and observations is only consistent in 
the broad energy range in terms of general trends, and our model does not agree
 with the observations in detail, e.g., there are the two declining regions 
around $10^5$ GeV and $10^9$ GeV in observations, which are not shown in the 
modeling results. On the other hand, the 
simplicity and analytical nature of our model allows for analytical 
calculations, and there is no need for much tuning factors in our model,
which only requires power law fitting of B/C ratio observations and a few basic 
parameters for the Galaxy, e.g., the half thickness $h$, gas number density
$n_\txt{d}$, of the galactic disk, and the spallation cross-section  
$\sigma_{\txt{C} \shortto \txt{B}}$. In 
addition, the modeling of the dipole anisotropy amplitude $\delta$ shows to varying 
with the CRs energy per nucleon, but observational data of $\delta$ vary with the 
total energy. We ignore the difference between the total energy and the energy per 
nucleon, considering the fact that the observed CRs are mostly protons. In addition, 
The fitting of the observational data of B/C ratio are in the energy range from tens of 
GeV/n to several TeV/n, but to obtain the modeling results of the dipole anisotropy 
amplitude, the fitting results of B/C ratio are used in the energy range from tens of GeV 
to hundreds of EeV. These could all introduce differences between modeling results 
and observations.

\begin{acknowledgments}
We were partly supported by grants NNSFC 42374190, NNSFC 42074206, NNSFC 42374189, 
and NNSFC 42150105. We were also partly supported by the Shenzhen Science and 
Technology Program under Grant No. JCYJ20210324132812029. The work were supported by 
the National Key R\&D program of China (No. 2021YFA0718600 and No. 2022YFA1604600), 
and by Shenzhen Key Laboratory Launching Project (No. ZDSYS20210702140800001). The 
work was supported by the Strategic Priority Research Program of Chinese Academy of 
Sciences, Grant No. XDB 41000000. We appreciate the availability of the B/C ratio 
data and the CR dipole anisotropy amplitude data at the Cosmic Ray Database Website 
(\url{https://lpsc.in2p3.fr/crdb}).
\end{acknowledgments}

%%%%%%%%%%%%%%%%%%%%%%%%%%%%%%%%%%%%%%%%%%%%%%%%%%%%%%%%%%%%%%%%%%%%%%%%%%%%%%%%%%%%%

%%%%%%%%%%%%%%%%%%%%%%%%%%%%%%%%%%%%%%%%%%%%%%%%%%%%%%%%%%%%%%%%%%%%%%%%%%%%%%%%%%%%%

%%%%%%%%%%%%%%%%%%%%%%%%% Figure & Table %%%%%%%%%%%%%%%%%%%%%%%%%%%%%%%%%%%%%%%%%%%%

\clearpage

\clearpage

\clearpage
\begin{table}
\caption{Experiments used for B/C data}
\label{tab:B/C}
\centering
\begin{tabular}{cc|cc|cc}
\hline\hline
Experiments & Symbols & Experiments & Symbols & Experiments & Symbols\\
\hline
ACE & $\textcolor{red}{\bullet}$  & CRN-Spacelab2 & $\textcolor{red}{\blacksquare}$ & ISEE3-HKH & $\textcolor{red}{\blacktriangle}$\\
ASM01 & $\textcolor{blue}{\bullet}$ & DAMPE & $\textcolor{blue}{\blacksquare}$ & OGO1 & $\textcolor{blue}{\blacktriangle}$\\
ASM02 & $\textcolor{green}{\bullet}$ & Gemini11 & $\textcolor{green}{\blacksquare}$ & PAMELA & $\textcolor{green}{\blacktriangle}$\\
ATIC02 & $\textcolor{cyan}{\bullet}$ & HEAO3-C2 & $\textcolor{cyan}{\blacksquare}$ & RICH-I & $\textcolor{cyan}{\blacktriangle}$\\
Balloon & $\textcolor{magenta}{\bullet}$ & IMP5 & $\textcolor{magenta}{\blacksquare}$ & TRACER & $\textcolor{magenta}{\blacktriangle}$\\
CALET & $\textcolor{yellow}{\bullet}$ & IMP7 & $\textcolor{yellow}{\blacksquare}$ & Ulysses-HET & $\textcolor{yellow}{\blacktriangle}$\\
CREAM-I & $\textcolor{gray}{\bullet}$ & IMP7 & $\textcolor{gray}{\blacksquare}$ & Voyager & $\textcolor{gray}{\blacktriangle}$\\
\hline
\end{tabular}
\end{table}

\clearpage
\begin{table}
\caption{Experiments used for the data of dipole anisotropy amplitude}
\label{tab:anisotropy}
\centering
\begin{tabular}{cc|cc|cc}
\hline\hline
Experiments & Symbols & Experiments & Symbols & Experiments & Symbols\\
\hline
ARGO & $\textcolor{red}{\bigstar}$ & IceCube & $\textcolor{yellow}{\blacklozenge}$ & Mt. Norikura & $\textcolor{cyan}{\blacktriangleleft}$\\
Artyomovsk & $\textcolor{blue}{\bigstar}$ & IceTop & $\textcolor{gray}{\blacklozenge}$ & Musala & $\textcolor{magenta}{\blacktriangleleft}$\\
Auger & $\textcolor{green}{\bigstar}$ & KASCADE & $\textcolor{red}{\blacktriangledown}$ & Nagoya & $\textcolor{yellow}{\blacktriangleleft}$\\
Baksan & $\textcolor{cyan}{\bigstar}$ & Kamiokande & $\textcolor{blue}{\blacktriangledown}$ & Plateau Rosa & $\textcolor{gray}{\blacktriangleleft}$\\
Bolivia & $\textcolor{magenta}{\bigstar}$ & LHAASO & $\textcolor{green}{\blacktriangledown}$ & Poatina & $\textcolor{red}{\times}$\\
Budapest & $\textcolor{yellow}{\bigstar}$ & Liapootah & $\textcolor{cyan}{\blacktriangledown}$ & Sakashita & $\textcolor{blue}{\times}$\\
Carpet & $\textcolor{gray}{\bigstar}$ & London & $\textcolor{magenta}{\blacktriangledown}$ & Socorro & $\textcolor{green}{\times}$\\
EAS-TOP & $\textcolor{red}{\blacklozenge}$ & MACRO & $\textcolor{yellow}{\blacktriangledown}$ & Super-Kamiokande & $\textcolor{cyan}{\times}$\\
EmbudoCave & $\textcolor{blue}{\blacklozenge}$ & Matsushiro & $\textcolor{gray}{\blacktriangledown}$ & Telescope Array & $\textcolor{magenta}{\times}$\\
HAWC & $\textcolor{green}{\blacklozenge}$ & Mayflower & $\textcolor{red}{\blacktriangleleft}$ & Tibet & $\textcolor{yellow}{\times}$\\
Hobart & $\textcolor{cyan}{\blacklozenge}$ & Milagro & $\textcolor{blue}{\blacktriangleleft}$ & Yakutsk & $\textcolor{gray}{\times}$\\
Honkong & $\textcolor{magenta}{\blacklozenge}$ & Misato & $\textcolor{green}{\blacktriangleleft}$ &  & \\
\hline
\end{tabular}
\end{table}

\end{document}